\documentclass[12pt]{article}
\usepackage{graphicx}
\usepackage{wrapfig}
\usepackage{amssymb}
\usepackage{cite}

\usepackage[a4paper,top=1.75cm,bottom=1.75cm,left=2.5cm,right=2.5cm,bindingoffset=0mm]{geometry}

\usepackage{xcolor}
\usepackage{hyperref}

\hypersetup{
	colorlinks,
	citecolor=blue,
	linkcolor=blue,
	urlcolor=blue
}

\newcommand\pubdate{\today}

\def\Title#1{\begin{center} {\Large #1 } \end{center}}
\def\Author#1{\begin{center}{ \sc #1} \end{center}}
\def\Address#1{\begin{center}{ \it #1} \end{center}}

\newcommand\pubblock{\rightline{\begin{tabular}{l}  \\ 
         \pubdate  \end{tabular}}}
\newenvironment{Abstract}{\begin{quotation}  }{\end{quotation}}
\newenvironment{Presented}{\begin{quotation} \begin{center} 
             PRESENTED AT\end{center}\bigskip 
      \begin{center}\begin{large}}{\end{large}\end{center} \end{quotation}}

\begin{document}
\begin{titlepage}
 \pubblock
\vfill
\Title{\Huge QCD effective charges from low-energy \\[0.2cm] neutrino structure functions}
\vfill
\Author{Tanjona R. Rabemananjara}
\Address{Department of Physics and Astronomy, Vrije Universiteit, NL-1081 HV Amsterdam\\[0.1cm]
  Nikhef Theory Group, Science Park 105, 1098 XG Amsterdam, The Netherlands\\[0.1cm]}
\vfill
\begin{Abstract}
We present a new perspective on the study of the behavior of the strong coupling $\alpha_s(Q^2)$
-- the fundamental coupling underlying the interactions between quarks and gluons as described
by the Quantum Chromodynamics (QCD) -- in the low-energy infrared (IR) regime.
We rely on the NNSF$\nu$ determination of neutrino-nucleus structure functions
valid for all values of $Q^2$ from the photoproduction to the high-energy region
to define an effective charge following the the Gross-Llewellyn Smith (GLS) sum rule.
As a validation, our predictions for the low-energy QCD effective charge
are compared to experimental measurements provided by JLab.
\end{Abstract}
\vfill
\begin{Presented}
DIS2023: XXX International Workshop on Deep-Inelastic Scattering and
Related Subjects, \\
Michigan State University, USA, 27-31 March 2023 \\
     \includegraphics[width=9cm]{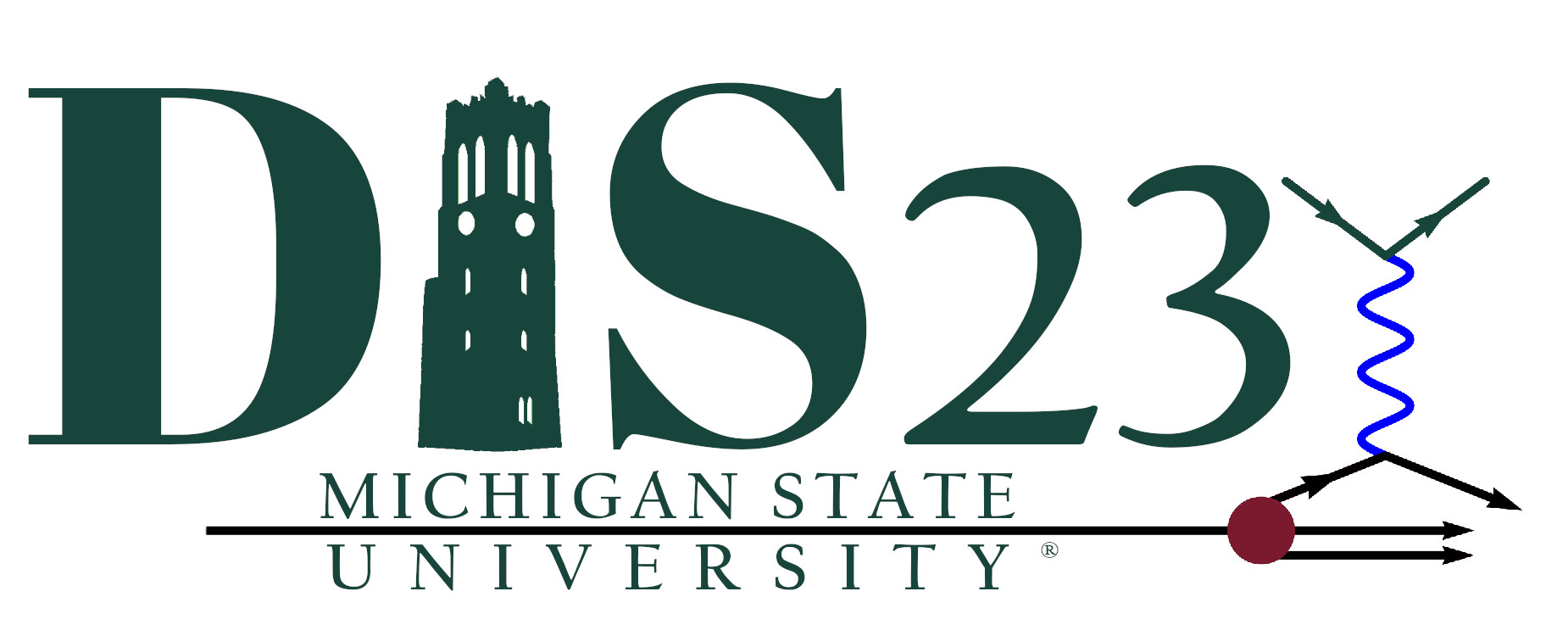}
\end{Presented}
\vfill
\end{titlepage}


\paragraph{Introduction.}
The study of (anti-)neutrino-nucleus interactions plays a crucial role in the interpretation
of ongoing and future neutrino experiments which ultimately will also help improve our general 
understanding of the
strong interactions as described by Quantum Chromodynamics (or QCD in short). Different types of
interactions occur depending on the neutrino energies $E_\nu$ probed. The one of particular relevance
to QCD is inelastic neutrino scattering, which occurs at energies above the resonance region,
for $E_\nu \gtrsim\mathcal{O}(10)~\rm{GeV}$ and when the invariant mass of the final states satisfies
$W \gtrsim 2~\rm{GeV}$. In such a regime, the inelastic neutrino scattering is composed of nonperturbative 
and perturbative regimes referred to as shallow-inelastic scattering (SIS) and deep-inelastic scattering
(DIS), respectively.

The main observables of interest in neutrino inelastic scattering are the differential cross-sections
which are expressed directly as linear combinations of structure functions 
$F_{i,A}^{\nu / \bar{\nu}} (x, Q^2)$ with $x$ the Bjorken variable, $Q^2$ the momentum transfer, and
$A$ the atomic mass number of the proton/nuclear target.
In the DIS regime, the neutrino structure functions are factorized as a convolution between the
parton distribution functions (PDFs) and hard-partonic cross-sections. The latter are calculable
to high order in perturbation theory while the former have to be extracted from experimental
data. On the other hand, in the SIS regime in which nonperturbative effects dominate, theoretical
predictions of neutrino structure functions do not admit a factorised
expressions in terms of PDFs. Various
theoretical frameworks have been developed to model these low-$Q^2$ neutrino structure functions,
e.g.~\cite{Yang:1998zb}, but all of them present limitations.

In~\cite{Candido:2023utz} we presented the first determination of neutrino-nucleus structure 
functions and their associated uncertainties that is valid across the entire range of $Q^2$ relevant 
for neutrino phenomenology, dubbed NNSF$\nu$. The general strategy comes down to dividing the $Q^2$ range 
into three distinct but interconnected regions. These regions refer respectively to the low-, intermediate-, 
and large-momentum transfers. At low momentum transfers $Q^2 \lesssim Q^2_{\rm dat}$ in which
nonperturbative effects occur, we parametrize the structure functions in terms of neural networks
(NN) based on the information provided by experimental measurements following
the NNPDF approach~\cite{AbdulKhalek:2022fyi}. In the intermediate momentum
transfer regions, $Q^2_{\rm dat} \lesssim Q \lesssim Q^2_{\rm thr}$, the NN is fitted to the DIS
predictions for convergence. And finally at large momentum transfers, $Q \gtrsim Q^2_{\rm dat}$,
the NN predictions are replaced by the pure DIS perturbative computations.

Such a framework allows us to provide more reliable predictions of the low-energy neutrino structure functions --
we refer the reader to~\cite{Candido:2023utz} for more details.
The NNSF$\nu$ enables the robust, model-independent evaluation of inclusive inelastic
neutrino cross-sections for energies from a few tens of GeV up to the several EeV
relevant for astroparticle physics~\cite{Bertone:2018dse}, and
in particular fully covers the
kinematics of present~\cite{FASER:2023zcr,snd} and future~\cite{Feng:2022inv,Anchordoqui:2021ghd} LHC neutrino experiments.

Aside from its relevance in studying neutrino physics, the NNSF$\nu$ framework may also potentially
be used as a tool to strengthen our understanding of the nonperturbative regions of QCD owing to its 
predictions in the low-energy regime. It is commonly understood that studying the theory of the strong
interactions in the Infrared (IR) regime is necessary to understand both high-energy and hadronic
phenomena therefore providing sensitivity to a variety of Beyond the Standard Model (BSM) scenarios.
One aspect that deserves a closer look in studying long-range QCD dynamics is the behavior of the
strong coupling $\alpha_s$ due to its special property as an expansion parameter for first-principle 
calculations. In the perturbative regime, the uncertainties in the value of $\alpha_s$ is known to
the sub-percent level ($\Delta \alpha_s / \alpha_s = 0.85\%$,~\cite{Deur:2016tte,Deur:2023dzc}). At 
low-$Q^2$, however, its determination is subject to large uncertainties mainly due to the lack of 
theoretical frameworks that can correctly accommodate for the nonperturbative effects.

A number of approaches have been explored in the literature to study the coupling in the nonperturbative
regime including lattice QCD or the Anti-de-Sitter/Conformal Field Theory (AdS/CFT) duality implemented
using QCD's light-front quantization. In the following, we use the Grunberg's effective charge approach
defined from the Gross-Llewellyn Smith sum rule sum rule. From perturbative QCD, the effective coupling
charge can be calculated from the perturbative series of an observable -- usually defined in terms of the
sum rules -- truncated to its first order in $\alpha_s$. The reason for such a truncation is related to
the scheme as at leading order the observable is independent of the renormalization scheme (RS). One of
the main advantages of the effective charge w.r.t. different approaches is that there are several
experiments that measure the effective coupling $\alpha_s^{\rm eff}(Q^2)$ to compare the theoretical
computations to.

Here first we briefly review the Gross-Llewellyn Smith sum rule
and verify that it is satisfied using the neutrino structure function predictions from the NNSF$\nu$
determination. We then use the NNSF$\nu$ framework to compute the effective charge defined from the
sum rule and compare the results to experimental measurements extracted at 
JLab~\cite{Deur:2008ej,ResonanceSpinStructure:2008ceg,Deur:2014vea,Deur:2021klh}.

\paragraph{The Gross-Llewellyn Smith sum rule.}
The neutrino structure function $x F_3^{\nu N}$ must satisfy the Gross-Llewellyn Smith (GLS) 
sum rule~\cite{Gross:1969jf} in which its unsubtracted dispersion relation has to be equal to the
number of valence quarks inside the nucleon $N$. Such a dispersion relation could also be
extended to the neutrino-nucleus interactions in which the GLS sum rule writes as follows:
\begin{equation}
	\mathrm{GLS}(Q^2, A) \equiv \frac{1}{2} \int_0^1 d x \left(F_3^{\nu A} + F_3^{\bar{\nu} A}\right) 
	\left(x, Q^2\right)= 3\left(1+\sum_{k=1}^3\left(\frac{\alpha_s\left(Q^2\right)}{\pi}\right)^k c_k
	\left(n_f\right)\right) + \frac{\Delta^{\rm HT}}{Q^2},
	\label{eq:gls-sum-rule}
\end{equation}
where $n_f$ is the number of active flavors at the scale $Q^2$. The terms inside the
parentheses on the right-hand side represent the perturbative contribution to the leading-twist
part whose coefficients $c_k$ have been computed up to $\mathcal{O}(\alpha_s^4)$. The
$\Delta^{\rm HT}$-term instead represents the power suppressed non-perturbative corrections,
see~\cite{Huang:2021kzc} for a recent review. Notice that the form of the perturbative part in 
Eq.~(\ref{eq:gls-sum-rule}) is convenient because, as opposed to many observables in pQCD, 
it does not depend both on $x$ and on the mass number $A$.

The low-energy experimental data from which the NNSF$\nu$ neutrino structure functions were determined
do not provide measurements in the low-$x$ region, and therefore the evaluation of
Eq.~(\ref{eq:gls-sum-rule}) largely depends on the modeling of the small-$x$ extrapolation region.
In our predictions, the behavior at small-$x$ is inferred from the medium- large-$x$ regions via
the preprocessing factor $x^{1-\alpha_i}$ whose exponents are fitted to the data. In addition,
due to the large uncertainties governing the small-$x$ region, we have to truncate the integration at
some $x_{\rm min}$ value. The truncated sum rule should however converge to the pQCD predictions
in the limit $x \to 0$. The truncated sum rule writes as
\begin{equation}
	\mathrm{GLS}(Q^2, A, x_{\rm min}) \equiv \frac{1}{2} \int_{x_{\rm min}}^1 d x 
	\left(F_3^{\nu A} + F_3^{\bar{\nu} A}\right) \left(x, Q^2\right),
	\label{eq:truncated-gls-sum-rule}
\end{equation}
for different values of the lower integration limit $x_{\rm min}$ and different nucleon/nuclear
targets.

\begin{figure}[!tb]
	\centering
	\includegraphics[width=0.495\linewidth]{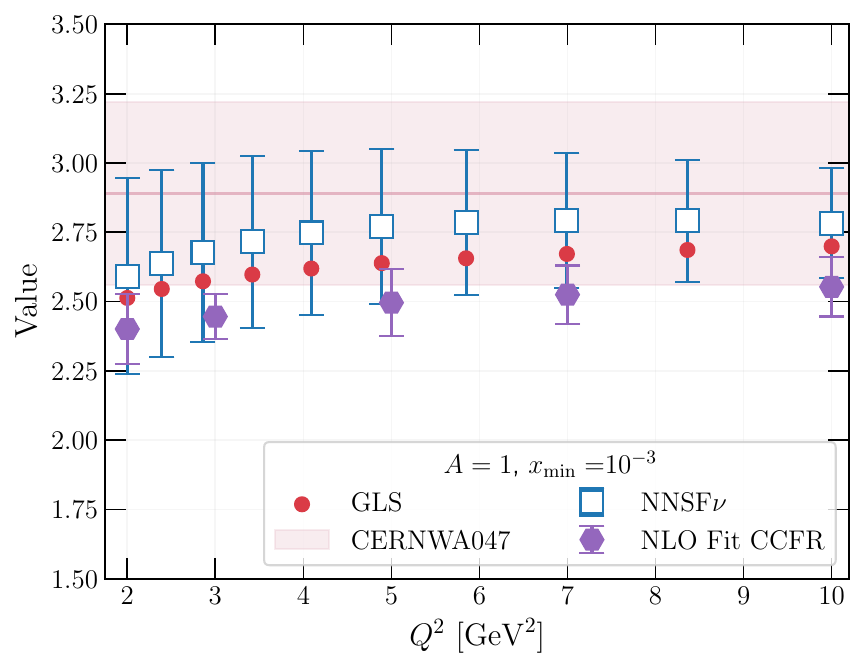}
	\includegraphics[width=0.495\linewidth]{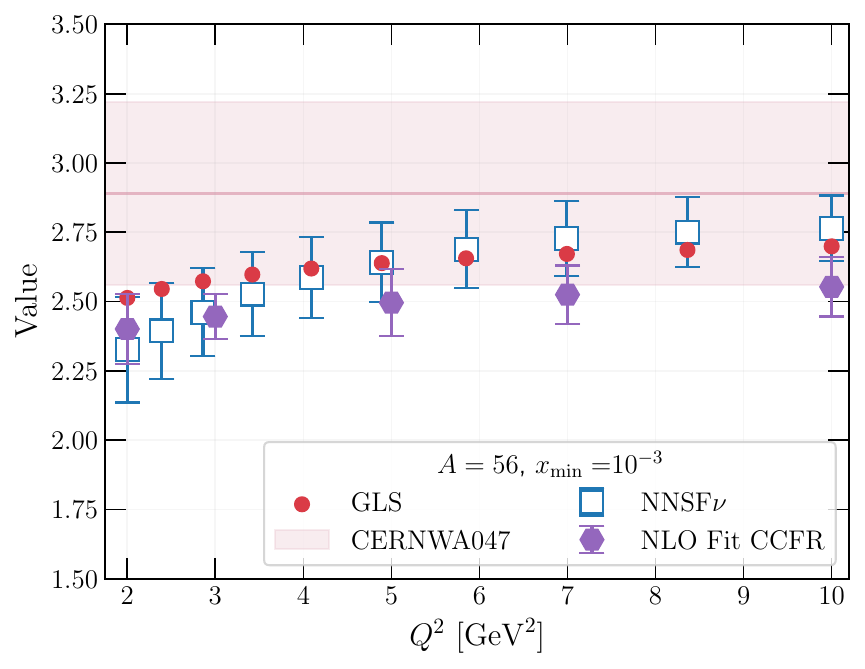}
	\includegraphics[width=0.495\linewidth]{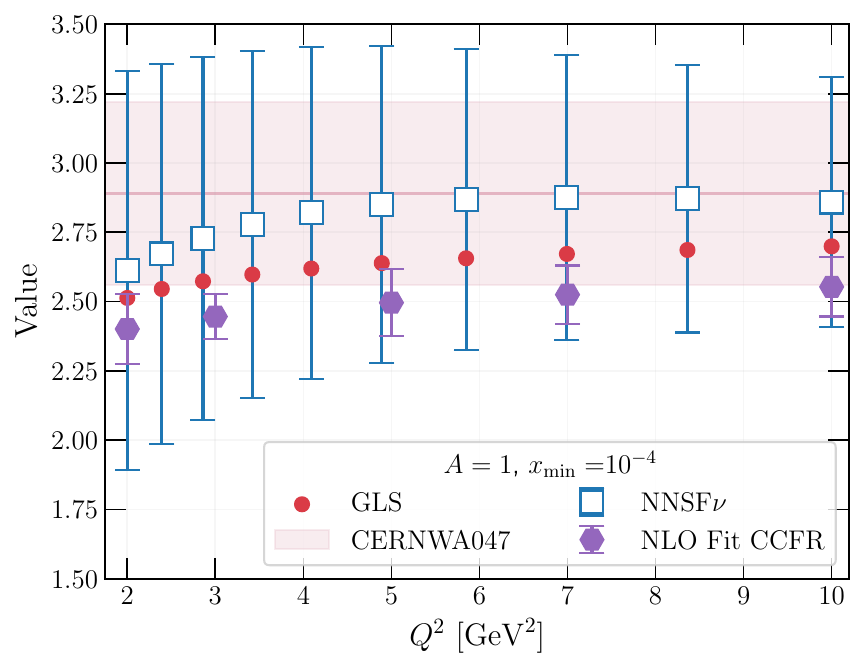}
	\includegraphics[width=0.495\linewidth]{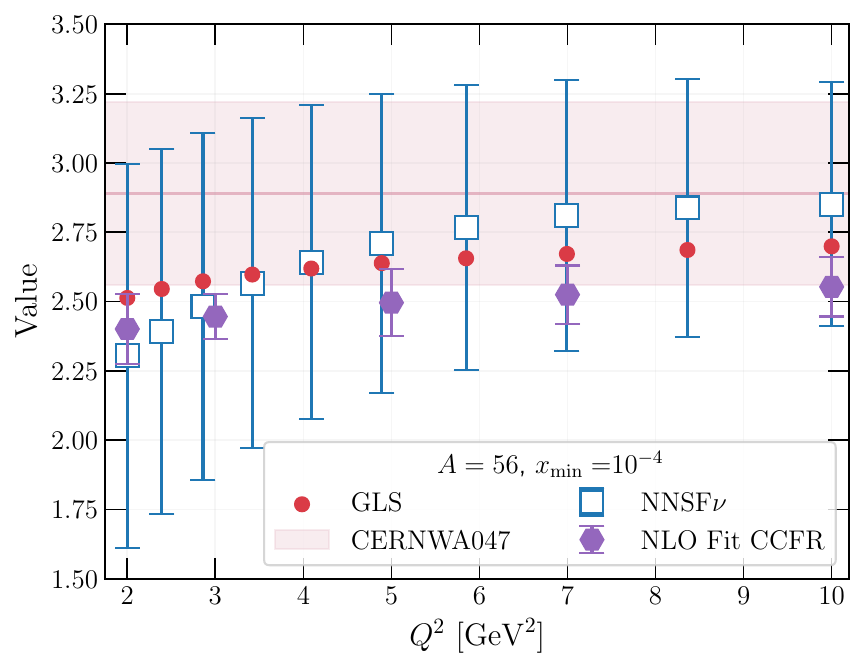}
	\caption{
	The GLS sum rule evaluated as a function of $Q^2$. The NNSF$\nu$ predictions are compared
	with the NLO fit to the CCFR data (purple hexagons), the CERN-WA-047 measurements (red band),
	and to the pure QCD predictions (red dots). The results are displayed for different lower 
	integration limits $x_{\rm min}=10^{-3}$ (top) and of $x_{\rm min}=10^{-4}$ (bottom panels),
	and for different atomic mass numbers $A=1$ (left) and for $A=56$ (right panels).
	}    
	\label{fig:GLS-sumrule}
\end{figure}

In Fig.~\ref{fig:GLS-sumrule} we display the results of computing the truncated GLS sum rule in 
Eq.~(\ref{eq:truncated-gls-sum-rule})
using our NNSF$\nu$ predictions. The results are shown for different lower integration limits
$x_{\rm min} = 10^{-3}, 10^{-4}$ and for different nuclei $A=1, 56$. For reference, we compare
the NNSF$\nu$ calculations with the NLO fit to the CCFR data~\cite{Kataev:1994ty}, 
the CERN-WA-047 measurements~\cite{Bolognese:1982zd},
and to the pure exact QCD predictions. All the results, except for the NNSF$\nu$, are always the
same in all the panels since they are independent of both $x$ and $A$. In the case of the QCD
predictions, the $Q^2$ dependence is entirely dictated by the running of the strong coupling $\alpha_s(Q^2)$.
As in the previous section, the error bars on the NNSF$\nu$ represent the 68\% confidence level intervals
from the $N_{\rm rep} = 200$ replicas fit.

Based on these comparisons, we can conclude that there is in general  good agreements between
the different results. In particular, the NNSF$\nu$ and pure QCD predictions are in prefect agreement
when the lower integration limit is taken to be $x_{\rm min} = 10^{-3}$. Even more remarkably, the
slope of the GLS sum rule, which in the the QCD computation is purely dictated by the running of
the strong coupling $\alpha_s(Q^2)$, is correctly reproduced by the NNSF$\nu$ predictions. The agreement
in central values slightly worsens when the lower integration limit is lowered down to $x_{\rm min} = 10^{-4}$.
Such a deterioration can also be seen in the increase of the uncertainties. As alluded earlier, such
a behavior is expected due to the fact that NNSF$\nu$ does not have direct experimental constraints below
$x \approx 10^{-3}$. Notice that the observations above hold for the different nuclei considered.

\paragraph{QCD effective charges.}
In order to fully understand the short- and long-range interactions, knowing the strong coupling $\alpha_s$ in
the nonperturbative domain (or equivalently in the IR regime) is crucial. Further arguments can be
put forth that knowing the IR-behavior of $\alpha_s$ is necessary to fully understand the mechanism for dynamical
chiral symmetry breaking~\cite{Deur:2016tte, Deur:2023dzc}.
However, studying the strong coupling in the IR domain is very challenging since standard perturbation theory 
cannot be used. In the following section, we explore an attempt to extend the perturbative domain using our NNSF$\nu$ 
framework to provide predictions for the low-energy strong coupling $\alpha_s (Q^2)$.

In the framework of perturbative QCD, the strong coupling -- which at leading order can be approximated as 
$\alpha_s(Q^2) = 4 \pi / \beta_0 \ln(Q^2 / \Lambda^2)$-- predicts a diverging behavior at the Landau Pole
when $Q^2 \to \Lambda^2$.  Such a diverging nature is not an inconsistency of 
perturbative computations per se since the pole is located in a region way beyond the ranges of validity 
of perturbative QCD. Instead, the origin of such a divergence is the absence of nonperturbative terms in
the series that cannot be captured by high order perturbative approximations. That is, the Landau singularity
cannot be cured by simply adding more terms to the perturbative expansion. The Landau Pole however is unphysical
(with the value of $\Lambda^2$ defined by the renormalization scheme) and this is supported by the fact that 
observables measured in the domain $Q^2 < \Lambda^2$ display no sign of discontinuity or unphysical behavior.

Several approaches have been explored to study the low-energy running of the coupling, each with its advantages,
justifications, and caveats. A prominent approach based on Grunberg's effective charge approach -- that we
attempt to pursue here -- provides a definition of the coupling that behaves as $\alpha_s^{\rm pQCD}$ at large-$Q^2$
but remains finite at small values of $Q^2$. Since the regime is extended down to small-$Q^2$, the effective
charge incorporates nonperturbative contributions that appear as higher-twist. Such an effective charge
is explicitly defined in terms of physical observables that can be computed in the perturbative QCD domain. An
example of such observable that has been very well studied in the literature is the effective charge
$\alpha_s^{\rm Bj} (Q^2)$ defined from the polarized Bjorken sum rule~\cite{PhysRev.148.1467, Ayala:2018ulm}. 
Such an observable has important advantages in that it has a simple perturbative series and is a non-singlet 
quantity implying that some $\Delta$-resonance contributions cancel out.

In the following study, we use the effective charge $\alpha_s^{\rm GLS} (Q^2)$ defined from the GLS sum rule
introduced above. Following the Grunberg's scheme, the definition of the effective charge $\alpha_s^{\rm GLS} (Q^2)$
which follows from the leading order of Eq.~(\ref{eq:gls-sum-rule}) writes as:
\begin{equation}
	\mathrm{GLS}(Q^2, A) \equiv 3 \left( 1 - \frac{\alpha_s^{\rm GLS} (Q^2, A)}{\pi} \right) \Longleftrightarrow
	\alpha_s^{\rm GLS} (Q^2, A) = \pi \left( 1 - \frac{\mathrm{GLS}(Q^2, A)}{3} \right).
	\label{eq:gls-effective-charge}
\end{equation}
In the perturbative domain $\Lambda^2 \ll Q^2$, we expect the effective charges from the Bjorken and GLS
sum rules to be equivalent $\alpha_s^{\rm GLS} (Q^2)=\alpha_s^{\rm Bj} (Q^2)$ up to $\mathcal{O}(\alpha^2_{\overline{\rm MS}})$.
In addition, at zero momentum transfer we expect $\alpha_s^{\rm GLS} (Q^2=0)=\alpha_s^{\rm Bj} (Q^2=0)=\pi$. The latter
kinematic limits originate from the fact that cross-sections are finite quantities and when 
$Q^2 \to 0 \Rightarrow x=Q^2/(2M\nu) \to 0$, the support integrand in Eq.~(\ref{eq:gls-sum-rule}) must also
vanish; therefore we have the following relations:
\begin{equation}
	\lim\limits_{Q^2 \to 0} \mathrm{GLS}(Q^2, A) = 0 \Longleftrightarrow \alpha_s^{\rm GLS} (Q^2=0)=\pi.
	\label{eq:limits}
\end{equation}

It is important to emphasize that Eq.~(\ref{eq:gls-effective-charge}) is directly related to the right-hand side
of Eq.~(\ref{eq:gls-sum-rule}). We can see from this definition of the coupling that both the short-distance
effects -- those within the parentheses of Eq.~(\ref{eq:gls-sum-rule}) -- and the long-distance perturbative
QCD interactions -- represented by the $\Delta^{\rm HT}$ term-- are incorporated into the expression of 
the effective coupling $\alpha_s^{\rm GLS} (Q^2)$.

\begin{figure}[!tb]
	\centering
	\includegraphics[width=0.496\linewidth]{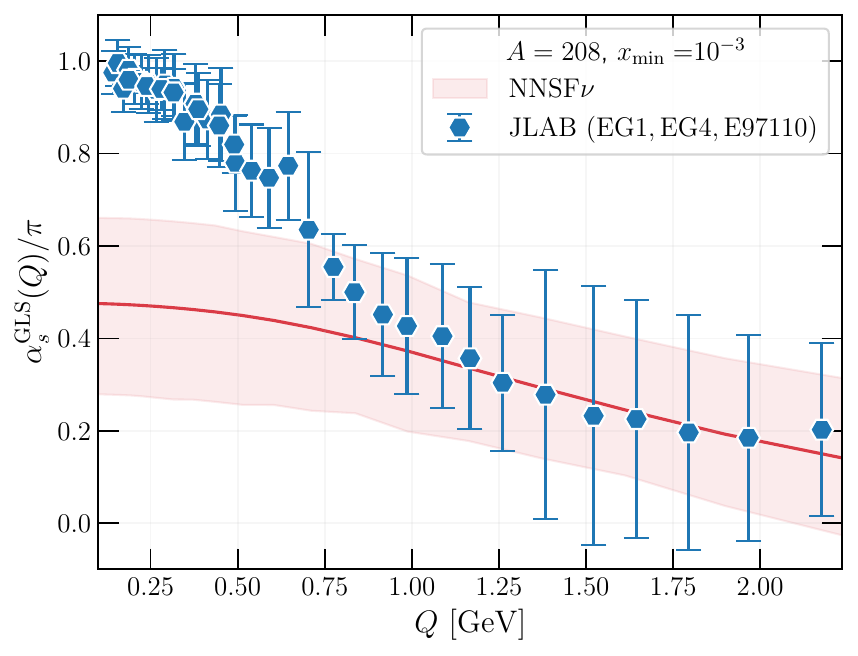}
	\includegraphics[width=0.496\linewidth]{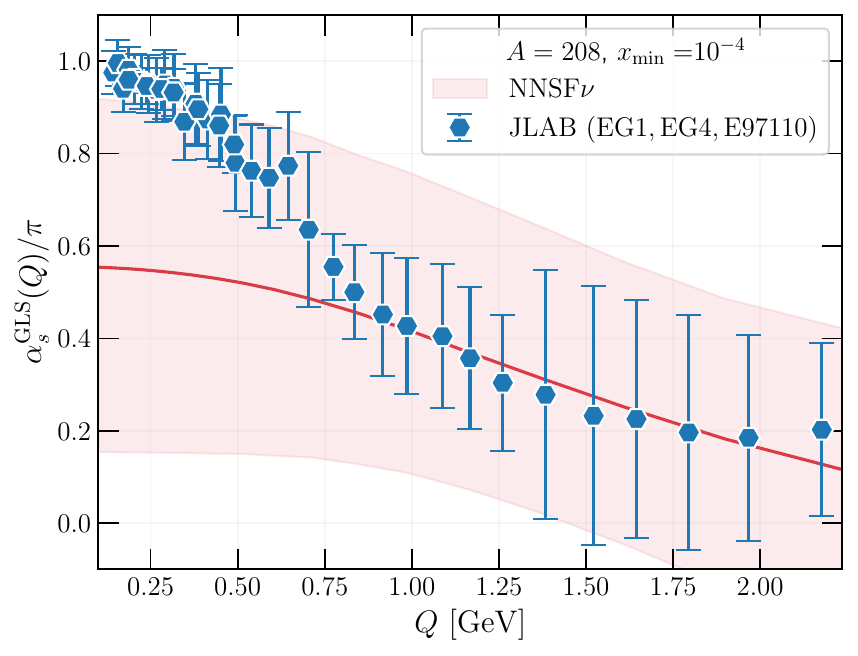}
	\caption{
		Effective charge defined in terms of the unpolarized GLS sum rule $\alpha_s^{\rm GLS} (Q)$
		computed from the NNSF$\nu$ predictions. The error bands represent the 68\% confidence level intervals. The predictions
		are shown for two lower integration limits $x_{\rm min} = 10^{-3}, 10^{-4}$. The results are compared to the 
		measurements of the Bjorken effective charge $\alpha_s^{\rm Bj}(Q)$ 
		from JLab~\cite{Deur:2008ej, ResonanceSpinStructure:2008ceg, Deur:2014vea, Deur:2021klh} 
		(blue hexagons) with the error bars representing the statistical and systematic uncertainties summed in quadrature.
	}    
	\label{fig:GLS-efcharge}
\end{figure}

Fig.~\ref{fig:GLS-efcharge} displays the effective coupling $\alpha_s^{\rm GLS} (Q)$ computed from
the NNSF$\nu$ predictions. As for the sum rules, the effective charge is truncated at some values $x_{\rm min}$
in order to not be influenced by the small-$x$ extrapolation region. The results are shown for $A=208$ and
for two different values of $x_{\rm min} = 10^{-3}, 10^{-4}$. Our predictions are compared to the
experimental measurements from JLab~\cite{Deur:2008ej, ResonanceSpinStructure:2008ceg, Deur:2014vea, Deur:2021klh} 
which measures the Bjorken effective charge $\alpha_s^{\rm Bj}(Q)$ using a polarized electron beam. Since the 
JLab results do not depend both on $x$ and on the atomic mass number $A$, the results are the same for all the panels.
This insensitivity of the results w.r.t the value of $A$ reflects the expectation that both the GLS and Bjorken sum rules
are related to the nucleon valence sum rules and therefore take the same values irrespective of the value
of $A$ entering the calculations.

Based on these comparisons, we can infer that the NNSF$\nu$ predictions and the JLab experimental measurements
agree very well down to $Q \sim 0.5~\rm{GeV}$. As $Q \to 0$ we can see that the effective coupling $\alpha_s^{\rm GLS}/\pi$
measured at JLab converges to $1$ as per the kinematic limit while our predictions converges to $\sim \langle 0.6 \rangle$.
Perhaps this result would slightly improve if the structure functions were forced to satisfy the sum rules during
the fit and if more experimental measurements were available to constrain the small-$x$ region. As before, the
decrease in the value of the lower integration limit $x_{\rm min}$ induces a significant increase in the
uncertainties.

\paragraph{Conclusions and outlook.}
In the first part of the manuscript, we reviewed a new framework -- referred to as NNSF$\nu$ -- for the determination 
of the neutrino-nucleus structure functions in the inelastic regime. In particular, we stressed on its
capabilities to provide predictions for low-energy neutrino interactions. As a verification of the methodology,
we compared the outcome of the computations of the GLS sum rule originating from such predictions with
measured experimental data to which we found very good agreement.

In the second part, we used the NNSF$\nu$ determination as a tool to understand the running of the coupling
$\alpha_s(Q^2)$ which encodes at the same time the perturbative dynamics at large momentum transfers and the
nonperturbative dynamics underlying the color confinement at small momentum transfers. The use of standard
perturbative computations to study the coupling at low-$Q^2$ yields erroneous results as it predicts a diverging
behavior due to the existence of an unphysical pole. Owing to the lack of theoretical formalism that
correctly accounts for the nonperturbative effects, studying the strong coupling in the IR regime is a
challenging task.

A prominent approach that resolves the ambiguity in defining the strong coupling in the nonperturbative
regime is the use of effective charges defined directly from a leading order perturbatively computable
observable. In our study we defined the effective charge $\alpha_s^{\rm GLS}(Q)$ from the GLS sum
rule which at large momentum transfers reproduces the perturbative computations and at low momentum
transfers is expected to converge to $\pi$. Our predictions yield comparable results to experimental
measurements -- accounting for the uncertainties -- down to $Q \sim 0.5~\rm{GeV}$. However, our
predictions do not fully satisfy the kinematic limit $\alpha_s(Q=0)/\pi = 1$ at zero momentum
transfers. This issue of convergence might be resolved by imposing the neutrino structure
functions to satisfy the sum rules during the fit. From this we conclude that further investigation
is needed in that direction in order to fully understand the $Q \sim 0$ behavior.

\paragraph{Acknowledgments.}
The author is grateful to Juan Rojo for the careful reading of the manuscript. T.~R. is supported by an 
ASDI (Accelerating Scientific Discoveries) grant from the Netherlands eScience Center.

\bibliographystyle{JHEP}
\bibliography{nnusf}
 
\end{document}